\documentstyle[twocolumn,aps,psfig,floats]{revtex}

\def\Sr14{Sr$_{14}$Cu$_{24}$O$_{41}$}
\def\Cax{Sr$_{14-x}$Ca$_{x}$Cu$_{24}$O$_{41}$}
\def\Cad{Sr$_{2}$Ca$_{12}$Cu$_{24}$O$_{41}$}
\def\SrCu2O3{SrCu$_2$O$_3$}
\def\Ca9{Sr$_{5}$Ca$_{9}$Cu$_{24}$O$_{41}$}

\begin{document}
\draft
\twocolumn[\hsize\textwidth\columnwidth\hsize\csname@twocolumnfalse\endcsname
%
% Title Page
%

\title{Absence of spin gap in the superconducting ladder
compound Sr$_2$Ca$_{12}$Cu$_{24}$O$_{41}$ }

\author{H. Mayaffre (1), P. Auban-Senzier (1), D. J\'erome (1),
D. Poilblanc (2), \\
C. Bourbonnais (3), U. Ammerahl (4)(5), G. Dhalenne (4) and A. Revcolevschi (4)
}

\address{(1) Laboratoire de Physique des Solides
 URA 002 (associ\'e au CNRS),\\
Universit\'e Paris-Sud, 91405 Orsay, France}

\address{(2) Laboratoire de Physique Quantique
UMR 5626 (associ\'e au CNRS), \\
Universit\'e Paul Sabatier, 31062 Toulouse, France}

\address{(3) Centre de Recherche en Physique du Solide,
D\'epartement de Physique, \\
Universit\'e de Sherbrooke, Sherbrooke, Qu\'ebec, Canada J1K2R1}

\address{(4) Laboratoire de Chimie du Solide,
UMR 446 (associ\'e au CNRS), \\
Universit\'e Paris-Sud, 91405 Orsay, France}

\address{(5) Physikalisches Institut, Universit\"{a}t zu K\"{o}ln,
Z\"{u}lpicher Strasse 77 D-50937 K\"{o}ln (Germany)}

\date{\today}
\maketitle

\begin{abstract}

Transport and  $^{63}$Cu-NMR, Knight shift and T$_{1}$, measurements
performed on the two-leg spin ladders of Sr$_2$Ca$_{12}$Cu$_{24}$O$_{41}$
single crystals show a collapse of the gap in ladder spin excitations when
superconductivity is stabilised under a pressure of 29~kbar. These results
support the prediction made with exact diagonalisation techniques 
in two-leg isotropic t-J ladders 
%@ BEGIN
of a transition 
%@ END
between a low-doping spin gap phase and a
gapless 1-D Tomonaga-Luttinger regime.

\end{abstract}

\vskip2pc]
\narrowtext

\section {Introduction}

The existence of superconductivity in two families of
materials where this property was not expected at first sight (the low
dimensional organic conductors and the high T$_c$ cuprates) has been a major
achievement of condensed matter research of the last two decades. The mechanism
of superconductivity for both classes of compounds is still under intense
debate but there is already a consensus about their low dimensional electronic
structure which may be the clue governing superconducting pairing
correlations. The recent finding of new superconducting copper-oxide
structures~\cite{Uehara96}
exhibiting one dimensional
features with both isolated CuO$_2$ chains and Cu$_2$O$_3$ ladders, that is,
pairs of CuO$_2$ chains linked by oxygen atoms between the coppers, has
profoundly
revived the interest for superconductivity in cuprates and 1-D materials.

The ladder compound to be discussed in this report is
Sr$_2$Ca$_{12}$Cu$_{24}$O$_{41}$ which derives from the parent compound
Sr$_{14}$Cu$_{24}$O$_{41}$ through covalent-calcium substitution. The structure
of Sr$_{14}$Cu$_{24}$O$_{41}$ displays together CuO$_2$ chains and Cu$_2$O$_3$
two-leg ladders parallel to the c-axis of the structure~\cite{McCarron88}.
%@ BEGIN
Note that other insulating 1-D materials like SrCu$_2$O$_3$
%~\cite{Hiroi91,Takano92}
contain only Cu$_2$O$_3$ ladders and no chain.
In contrast, the undoped parent
compound for the high T$_c$ cuprates exhibits a two dimensional (2-D)
CuO$_2$ layer structure.
In both systems, all copper sites belonging to the ladders or to the planes
are occupied by a spin 1/2 Cu$^{2+}$ ion.
However, while long range antiferromagnetism is stabilized at low
temperature in the 2-D spin system, the properties
of the spin ladder materials can be drastically different.
In two-leg ladders systems,
dominant AF coupling $J$ between the copper spins on the same rung
leads to the formation of a spin singlet on each rung.
%@ END
Consequently the
ground state of the whole ladder is a singlet spin state with a finite energy
needed to
excite a rung spin singlet to a spin triplet state. A spin gap situation is
obtained and can be seen via the exponential drop of the spin susceptibility
upon cooling down.

The existence of a spin gap in a spin-ladder
structure has been first proposed
theoretically~\cite{Dagotto92}
and found experimentally in several even-leg ladder copper oxide systems
(SrCu$_2$O$_3$~\cite{Azuma94,Dagotto96}, LaCuO$_{2.5}$~\cite{Hiroi95})
or organic materials~\cite{Imai96}.
The spin gap is expected to be quite robust to
various perturbations.
For example, it is predicted to be stable up to arbitrary small
magnetic coupling along the rungs of the ladders~\cite{SG_anisotropy} or
in the presence of a small inter-ladder coupling~\cite{Poilblanc94}.

The persistence of the spin gap upon light
hole doping has been established by various techniques (see e.g.
Refs.~\cite{Dagotto92,pairing,Poilblanc95,note-doping}). Even more exciting was
the suggestion that the spin gap leads to an attractive
interaction~\cite{Dagotto92,pairing} of holes on the same rung, 
hence providing dominant d-wave like superconducting pairing 
correlations~\cite{Hayward95,supercond_Hubbard} which could possibly 
materialize into a three-dimensional superconducting state at 
low temperature. In addition, the numerical investigation of the complete 
phase diagram~\cite{phase-diagram} of the two-leg isotropic $t-J$ ladder 
(in this model the motion of the Cu$^{3+}$ singlet holes are described by 
a hopping matrix element $t$) suggests the possibility of a transition
from the low doping spin gap phase (with the concomitant formation of
hole pairs) towards a gapless 1-D Tomonaga-Luttinger (TL) regime.
Although, the existence of such a transition is clear
at sufficiently large doping (where, like in
the non-interacting picture, the bonding band only becomes relevant),
there are some indications that such a TL phase might also be stable at
small doping and small $J/t$ ratio (typically around $J/t$~$\sim$~0.2
for a hole concentration of 0.2 hole per ladder-Cu).
%@ END

Some similarities between the superconductivity in Q-1-D organic
conductors~\cite{Jerome91}  and in ladder copper oxides have been noted
~\cite{Maekawa96} since, for both cases,
superconductivity
arises once a charge localized state is suppressed above a critical pressure.
The resemblance has to be taken "cum grano salis" since the band filling is
exactly one half in organic conductors whereas it is definitely different from
one half and possibly pressure-dependent, in the case of ladders. Furthermore,
the localization process is due to the existence of a spin density wave state
in organics~\cite{Jerome91} while it can be attributed to a charge density
wave state in ladder compounds.

Optical conductivity measurements of
Sr$_{14-x}$Ca$_{x}$Cu$_{24}$O$_{41}$ have shown that holes are transferred
to the
ladders from the chains upon increasing the Ca
concentration~\cite{Osafune97} . A copper
valency of
$\sim$2.2 is therefore reached in Sr$_{2}$Ca$_{12}$Cu$_{24}$O$_{41}$ (hole
density of 0.2 per ladder-Cu). What pressure might do is to further increase
the density of holes in ladder-Cu and/or possibly decrease the ratio $J/t$
and trigger the transition from the spin gap regime to
a TL regime with low-lying spin excitations.

The prediction of a coexistence, in the lightly doped regime, of
the spin gap with divergent superconducting correlations~\cite{Hayward95} is
very suggestive for experimentalists. Therefore, it is  a crucial
experimental test
to investigate
whether the finite spin gap persists in  spin
excitations of Sr$_{14-x}$Ca$_{x}$Cu$_{24}$O$_{41}$ when superconductivity of
Ca-substituted compounds is stabilized under pressure.
We present the preliminary
results of transport and $^{63}$Cu-NMR studies performed on a
Sr$_2$Ca$_{12}$Cu$_{24}$O$_{41}$ single crystal under pressure. We show, at
30~kbar, the
simultaneous onset of superconductivity below 5~K and the existence of
low-lying spin
excitations from the temperature dependence of the $^{63}$Cu Knight shift
belonging to
the ladders.

\section {Experimental results}

The two studies have been carried out on two samples
($(0.75\times 0.65 \times 1.18)$ mm$^3$ for transport experiments and
$(2.12 \times 0.9 \times1.12)$ mm$^3$
for NMR, along a, b and c-axes respectively) cut out from a slice of a
monodomain single
crystal, several centimeters long, grown by the travelling solvent floating
zone method
in an infrared image furnace under an oxygen pressure of 13~bar~\cite{Revco96}.

Transport data have been obtained by a conventional
four-contact AC lock-in
technique using a non- magnetic high pressure clamped cell. The 73~$\%$
increase of the
c-axis conductivity at 30~kbar is in fair agreement with the other single
crystal data
in the literature~\cite{Motoyama97,Akimitsu97}. Under 26~kbar,  $\rho _c$
reveals a
metal-like
temperature dependence, with a charge localisation beginning
around 30~K but no
sign of superconductivity above 1.8~K which was our lowest temperature. The
metal-like
behavior of $\rho _c$ is quite similar under 29~kbar (Fig. 1) but the weak
localization at low
temperature is stopped at 5~K by a sudden drop of the resistance which
compares rather well with the pressure data obtained on a
sample which was slightly less hole-doped~\cite{Akimitsu97}.
We believe that the
resistance drop can be confidently ascribed to the onset of a pressure-induced
superconducting ground state.

\begin{figure}[htb]
\begin{center}
\psfig{figure=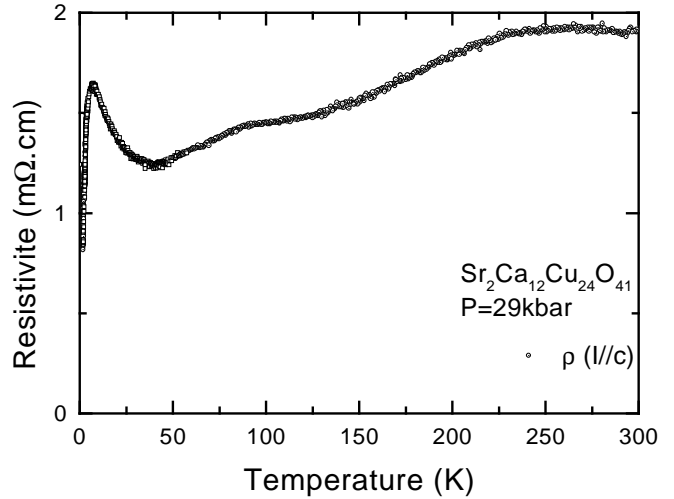,width=8.6truecm,angle=0}
\end{center}
\caption{Temperature dependence of the resistivity along the ladders under
29 kbar. The onset of superconductivity is at 5 K.}
\label{Fig1}
\end{figure}

The $^{63}$Cu-NMR experiments have been performed
in the same high pressure cell at an NMR frequency of 87.05~MHz. $^{63}$Cu-NMR
signals were obtained by recording the spin-echo amplitude of the central
transition (1/2, -1/2) of the copper nuclei pertaining to the ladders. A narrow
$^{63}$Cu-NMR signal has also been obtained by the usual Fourier transform
method. It has been attributed to $^{63}$Cu nuclei in the pressure cell
surrounding the sample. Since the Knight shift of copper metal is known to be
insensitive to pressure on the scale of Knight shift variations to be discussed
below~\cite{Benedek58} the narrow signal has been used as a
sensitive in-situ magnetic field marker.

Figure 2 presents the
temperature dependence of the $^{63}$Cu-NMR line position with H//b-axis under
30~kbar and also, in the same pressure cell, after releasing the pressure. The
central line position is affected by magnetic shift and by second order
quadrupolar shift. We estimate here the latter to be about 300~ppm. Moreover we
know that the variation of the quadrupolar frequency with temperature did not
exceed 10~\%~\cite{Caretta97}. The scattering of the data at low
temperature is due to an increase of the linewidth below 30~K. The magnetic
shift
consists usually of a temperature-independent orbital contribution, plus
the spin part which follows the uniform susceptibility and can possibly be
temperature and pressure-dependent.
The ambient pressure curve in Fig. 2 has been obtained by
subtraction of a constant contribution (1.36~\%) from the total  shift. The
ambient pressure data lead to K$_{orb}~\sim$~1.33~\% and $\Delta _{sp} \sim$
 250~K, given the expression for the T-dependent part~\cite{Troyer94}:
$$K_s(T) \sim \frac{1}{\sqrt{T}}\exp{\left( -\frac{\Delta _{sp}}{kT}\right),
}$$
at $kT \ll \Delta_{sp}$. Making the reasonable assumption of a
P-independent orbital
shift, and a negligeable variation of the quadrupolar contribution, the
30~kbar
data in Fig. 2 show only a weak pressure dependence of the spin part at high
temperature ($T > 150$~K) which is at variance with the drastic pressure
dependence of the spin part observed at low temperature. Fig. 2 reveals the
emergence of a zero temperature susceptibility which amounts to 53~\% of
the room
temperature value. Our data at ambient pressure provide a confirmation for the
depression of the spin gap upon doping, as previously announced in the Knight
shift study of a Sr$_5$Ca$_9$Cu$_{24}$O$_{41}$ single crystal
~\cite{Magishi97} and also from T$_1$ data in the whole series
corresponding to $0\leq x\leq 9$ ~\cite{Kumagai97}.  However, what is remarkable in
Fig. 2 is the existence of low lying spin excitations in the compound \Cad
under a
pressure of 30~kbar which is large enough to evidence the onset of
superconductivity at 5~K. First,the coincidence between the stabilization of a
superconducting ground state and the collapse of the ladder spin gap is a
strong argument in favor of superconductivity taking place on the ladders.

\begin{figure}[htb]
\begin{center}
\psfig{figure=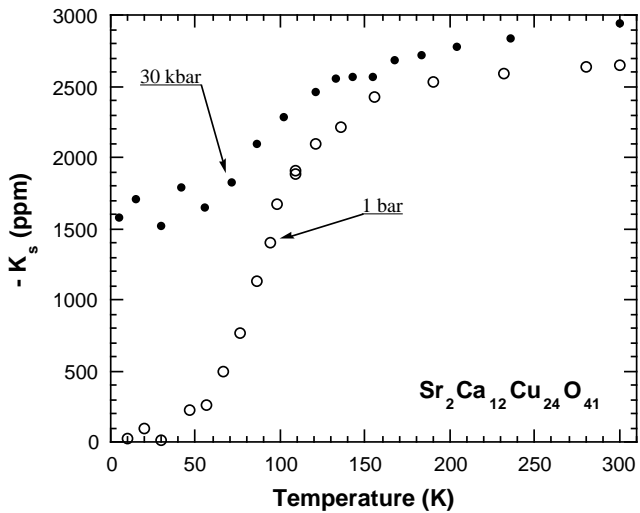,width=8.6truecm,angle=0}
\end{center}
\caption{Temperature dependence of the $^{63}$Cu-ladder nuclei Knight
shift. A spin gap $\Delta_{sp} \sim 250$ K is obtained from an activation
plot  below 150K. }
\label{Fig2}
\end{figure}

Comparing 1~bar and 29~kbar resistivity data we can infer
from the smaller increase of the
$\rho _a$/$\rho _c$ anisotropy on cooling at 29~kbar
that the confinement of the carriers along the ladders could correlate with
the size of the spin gap ~\cite{Orsay97}. This
finding suggests a picture of hole pairs being responsible for the conduction
within the ladders as long as the magnetic forces can provide the binding of
two holes on the same rung~\cite{Dagotto92}. The vanishing of the spin gap
could thus be responsible for the dissociation of the pairs, making in turn the
hopping of the transverse single particle easier.
Secondly, the behavior of the susceptibility under 30~kbar with
low-lying spin excitations  is at first sight  reminiscent of the situation
which prevails in
quasi 1-D organic conductors where the susceptibility is  temperature
dependent below 300~K,  but  noticeably independent of the behavior of the
charge
degrees of freedom (spin-charge separation in a 1-D chain), which have been
understood in terms of a correlated Tomonaga-Luttinger liquid~\cite{Jerome91}.

\begin{figure}[htb]
\begin{center}
\psfig{figure=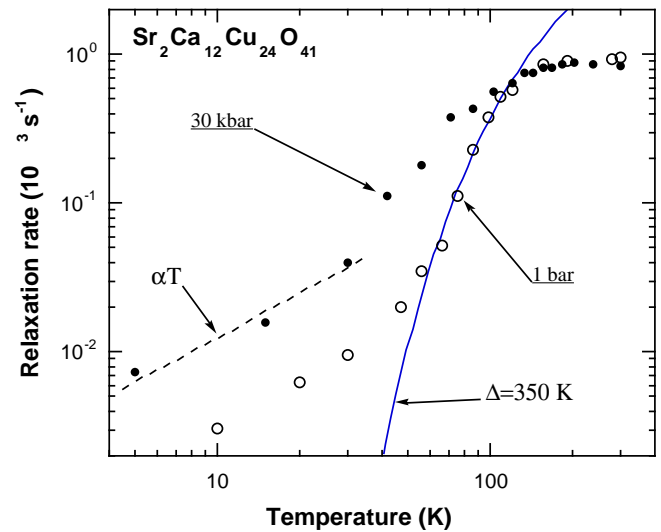,width=8.6truecm,angle=0}
\end{center}
\caption{Temperature dependence of the $^{63}$Cu-ladder relaxation rate
$T_1^{-1}$ at 1~bar and under 30~kbar.}
\label{Fig3}
\end{figure}

%@ NOTE: Paragraphe a revoir
The nuclear spin lattice relaxation has also been measured since this quantity
is known to be a very sensitive probe for the low lying spin excitations in
cuprates~\cite{Berthier96} and 1-D organic conductors
~\cite{Bourbonnais93}. The spin-lattice
relaxation with H//b was measured by the saturation-recovery technique and
$T_{1}$
was determined from the single time constant governing the magnetic
relaxation in
high field NMR of a quadrupolar nucleus such as $^{63}$Cu (I=3/2)
~\cite{Narath76}.
The exponential fit with a single time constant is excellent down to 70~K
but is not as good at low temperature and leads to a poorer $T_{1}$
determination.
The 1~bar and 30~kbar data for $\frac{1}{T_{1}} \  vs\  T$ in logarithmic
scales are
shown on
Fig.3. In the temperature domain 300-200~K, $T_{1}$ is both temperature and
pressure independent. At 1~bar, $T_{1}$ becomes activated below 120~K with an
activation
energy $\Delta '$ = 350~K. This behavior is in fair agreement with other
experiments on single crystals ~\cite{Magishi97}. This activated behavior
breaks down
below 40~K while under 30~kbar, the relaxation apparently displays quite a
different temperature
dependence.  We can attribute the non activated temperature dependence
to the collapsing of the spin gap  and the persistence even at low
temperature of populated
spin excitations modes contributing to the relaxation. At  high
temperature, however, the absence of significant temperature dependence
suggests a relaxation induced by antiferromagnetic
fluctuations in Heisenberg chains.
In this range of temperature,  such a relaxation channel should not be
sensitive to the presence or the
absence of a spin gap.

\section {Discussion}
%@ BEGIN
We finish this paper by discussing the above experimental
data in the context of existing theoretical work.
As stated above, such materials are expected to be accurately described
by a two-leg $t-J$ ladder which accounts for the strong nearest-neighbor
AF coupling $J$ between the spins and the hole delocalization $t$.
It is believed that, in these materials, interchain and intrachain
couplings are quite similar
(isotropic case) and comparable in magnitude to their values in the
2-D copper oxyde materials (typically $J/t\simeq 0.3 \ldots 0.4$ corresponding
to the strong coupling limit).
Exact diagonalisation techniques
applied to the two-leg $t-J$ ladder have proven, in the isotropic case,
the formation of hole pairs at low doping with the
concomitant formation of a spin gap~\cite{Dagotto96,pairing,Poilblanc95},
as initially suggested in the anisotropic limit ($J_\perp >> J_\parallel$).
A phase diagram for the isotropic $t-J$ ladder as a function
of $J/t$ and doping parameters has been established~\cite{phase-diagram}
and is schematically shown in Fig. 4. Close
to the half-filled band situation, i.e. with a small hole concentration,
the spin gapped phase (with a zero energy charge mode) is recovered
while a phase with a
single gapless spin and a single gapless charge mode is found to become stable
under hole doping~\cite{Poilblanc95}. Physically, the existence of such a
transition can easily be understood from a weakly-interacting band
picture~\cite{weak-coupling}.
Indeed, for hole density $n_h\ge 0.5$ the higher energy anti-bonding
band becomes unoccupied and one recovers a single band picture analogous
to the single chain case.
However, the line $n_h=0.5$ is expected to be quite singular since
Umklapp scattering characteristic of a half-filled band
likely leads to an insulating state~\cite{TMRice} and it is not yet clear
whether the transition line to the TL phase corresponds precisely to this
line. More interestingly, the possiblity of a similar cross-over
between the spin gapped phase towards a 1-D TL regime also exists at 
low doping for a sufficiently small $J/t$ ratio,
although it is difficult to estimate numerically the transition
line~\cite{TMRice} (indicated tentatively in Fig. 4 by a dotted line).

\begin{figure}[htb]
\begin{center}
\psfig{figure=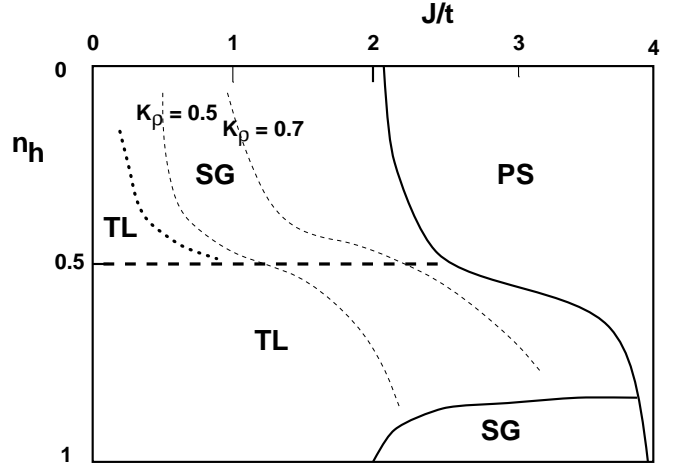,width=8.6truecm,angle=-90}
\end{center}
\caption{Schematic phase diagram of the two-leg t-J ladder as a function of
$J/t$ and hole density $n_h$ based on exact diagonalisations of small
clusters (see Ref.~\protect\cite{phase-diagram}). The thick dashed and dotted
lines separate the spin gapped (SG) and Tomonaga-Luttinger (TL) phases.
At large unphysical $J/t$ ratios phase separation (PS) occurs.   
Lines of constant $K_\rho$ (see text) are also indicated (thin dashed lines).  
}
\label{Fig4}
\end{figure}

We shall argue here that the experimental data of $T_{1}$, together
with the loss of susceptibility by a factor about 2 could possibly support
the existence of such a transition in the \Cax  materials,
the 1~bar and 29~kbar phases being ascribed to the spin gapped and TL 
phases respectively. From simple symmetry considerations~\cite{Poilblanc95},
in a ladder system one expects two spin excitations branches which could 
independently acquire a gap or be gapless.
The exponential decrease of the spin susceptibility at 1~bar (Fig. 2) is
characteristic of a full gap in the spin sector. On the other hand, the TL
phase exhibits one gapless mode and one mode with a gap.
The susceptibility data at 29~kbar are consistent with this scenario.
At sufficiently high temperatures both modes are expected to contribute.
However, the gapped spin mode is depopulated on decreasing temperature and
a pseudo-gap feature can be seen. At the lowest temperatures half of the
spin degrees of freedom contribute to the static susceptibility as
portrayed by the temperature variation of $K_s$ in Fig. 2. 

It is worth
looking at the temperature profile of the relaxation rate that would be
predicted in the above scenario of the
$t-J$ ladder model. First, since for $T~>~150$K, $T_1^{-1}$   is found to
be temperature independent indicating that antiferromagnetic  chain-like
spin correlations dominate the relaxation.  This suggests that  the
contribution coming from  uniform  spin correlations to
$T_1^{-1}$ is sufficiently small to be safely ignored for all temperatures
and pressures  of interest~\cite{Bourbonnais93}. In the spin gap phase at
1~bar, where both branches of spin excitations are frozen,  $T_1^{-1}$ and
$K_s$ are  thermally activated with slightly different activation
energies~\cite{Magishi97}. At high pressure, however,  the restoration of a
phase with a gapless spin and  a gapless charge modes will promote a new
channel of relaxation  of the TL type.  This is well known to introduce  a
power law component in the temperature dependence of  $T_1^{-1}$
which will read
$$
 T_1^{-1} = C_1 \exp(-\Delta'/k_BT)\  + \ C_2 T^{2K_\rho}
$$
in the low temperature domain,  where $C_{1,2}$ are positive constants.
Here $K_\rho$ stands as
the power law exponent of the antiferromagnetic LL response function ($\chi
\sim T^{-1 + 2 K_\rho}$)~\cite{phase-diagram}. Since in the  metallic phase
that is precursor to  superconductivity,
$2K_\rho$ should be close to unity, it follows that a non thermally activated
 component sould emerge for
$T_1^{-1}$  at sufficiently low temperature. This prediction seems to be
in qualitative agreement with  the $T_1^{-1}$  data given in Fig. 3. Indeed
one finds no
thermal activation for the relaxation and this, especially in the
temperature range  below 50~K where
$K_s$  becomes metallic and temperature independent.

In summary, transport and $^{63}$Cu-NMR measurements under pressure 
performed on the prototype \Cad  doped spin ladder have been reported
and analyzed in the context of the t--J ladder model. The drastically 
different behaviors observed at 1~bar and 29~kbar of the temperature 
dependence of the local static susceptibility and of the relaxation rate 
$1/T_1$ are attributed to the appearance of a zero energy spin mode at 29~kbar.
We argue that this phase can be identified to the gapless TL metallic phase
of the t-J ladder. Since, in the 29~kbar phase,
superconductivity sets in at low temperature, this suggests that 
superconductivity itself might be connected to such a transition and to 
the collapse of the spin gap. Recent numerical investigations of 
t--J~\cite{Riera97} and Hubbard~\cite{supercond_Hubbard} ladders found that
pairing correlations are maximized when the Fermi level lies in a 
maximum of the density of states, situation which might correspond to 
the transition discussed in this paper. 
Clearly more experimental and theoretical work are needed to clarify this
important issue. 

H. Mayaffre thanks the `Direction de la Recherche et  de la Technologie'
(DRET) for financial support. We acknowledge the technical support of M.
Nardone for high pressure experiments.
U. Ammerahl acknowledges support from DAAD in the
frame of the Procope program and the graduiertenkolleg of the Deutsche
Forschungsgemeinschaft.


\begin{references}

\bibitem{Uehara96}
M.~Uehara, T.~Nagata, J.~Akimitsu, H.~Takahashi, N.~M\^ori, and K.~Kinoshita,
{\em J. Phys. Soc. Japan.}, {\bf 65}, 2764 (1996).

\bibitem{McCarron88}
E.M.~Mc. Carron, M.A. Subramanian, J.C. Calabrese, and R.L. Harlow,
{\em Mater. Res. Bull.}, {\bf 23},1355 (1988);
T.~Siegrist, L.F. Schneemeyer, S.A. Sunshine, J.V. Waszczak, and R.S. Roth,
{\em Mater. Res. Bull.}, {\bf 23},1429 (1988).

\bibitem{Dagotto92}
E.~Dagotto, J.~Riera and D. J. Scalapino,
\newblock {\em Phys. Rev. B}, {\bf 45}, 5744 (1992); see also
H. J. Schulz, \newblock {\em Phys. Rev. B}, {\bf 34}, 6372 (1986);
E.~Dagotto and A.~Moreo,
\newblock {\em Phys. Rev. B}, {\bf 38}, 5087 (1988).

\bibitem{Azuma94}
M.~Azuma, Z.~Hiroi, M.~Takano, K.~Ishida, and Y.~Kitaoka,
{\em Phys. Rev. Lett.}, {\bf 73}, 3463 (1994).

\bibitem{Dagotto96}
E.~Dagotto and T.M. Rice,
{\em Science}, {\bf 271}, 618 (1996).

%\bibitem{Eccleston94}
%R.S. Eccleston, T.~Barnes, J.~Brody, and J.W. Johnson, {\em Phys. Rev.
%Lett.}, {\bf 73}, 2626 (1994).

\bibitem{Hiroi95}
Z.~Hiroi and M.~Takano, {\em Nature}, {\bf 377}, 41 (1995).

\bibitem{Imai96}
H.~Imai, T.~Inabe, T.~Otsuka, T.~Okuno, and K.~Anaga, {\em Phys. Rev. B},
{\bf 54}, R6838 (1996);
Spin gap features in the new $Cu_2(C_5H_{12}N_2)_2Cl_4$ material
are discussed in G.~Chaboussant, P.~Crowell, L.P.~L\'evy, O.~Piovesana,
A.~Madouri and D.~Mailly, {\em Phys. Rev. B}, {\bf 55}, 3046 (1997).

\bibitem{SG_anisotropy}
T.~Barnes, E.~Dagotto, J.~Riera and E.~Swanson,
{\em Phys. Rev. B}, {\bf 47}, 3196 (1993); see also S. Gopolan,
T. M. Rice and M.~Sigrist, {\em Phys. Rev. B}, {\bf 49}, 8901 (1994).

\bibitem{Poilblanc94}
D.~Poilblanc, H.~Tsunetsugu, and T.M. Rice, {\em Phys. Rev. B}, {\bf 50},
6511 (1994).

\bibitem{pairing}
M.~Sigrist, T.M. Rice, and F.C. Zhang, {\em Phys. Rev. B},  {\bf 49},12058
(1994);
H.~Tsunetsugu, M.~Troyer, and T.M. Rice, {\em Phys. Rev. B}, {\bf 49},16078
(1994).

\bibitem{Poilblanc95}
D.~Poilblanc, D.~J. Scalapino, and W.~Hanke, {\em Phys. Rev. B}, {\bf 52},
6796 (1995).

\bibitem{note-doping}
Doping effects in related 1-D models with spin gaps have been studied
previously. See e.g. M.~Ogata et al., \newblock {\em Phys. Rev. B}, {\bf
44}, 12083 (1991); M. Imada, {\em J. Phys. Soc. Jpn.} {\bf 60}, 1877
(1991).

\bibitem{Hayward95}
C.~Hayward, D.~Poilblanc, R.M.~Noack, D.J.~Scalapino, and W.~Hanke, {\em
Phys. Rev. Lett.}, {\bf 75}, 926 (1995).

\bibitem{supercond_Hubbard}
Enhanced d-wave pairing correlations have also been detected recently in
anisotropic Hubbard ladders by Density Matrix Renormalisation Group 
techniques and Monte Carlo simulations. See e.g. R.M.~Noack, N.~Bulut,
D.J.~Scalapino and M.G.~Zacher, cond-mat/9612165 preprint (1996).

\bibitem{phase-diagram}
C.~Hayward and D.~Poilblanc, {\em Phys. Rev. B}, {\bf 53}, 11721 (1996).
See also
H.~Tsunetsugu, M.~Troyer, and T.M. Rice, {\em Phys. Rev. B}, {\bf 51},
16456 (1995).

%\bibitem{Rice93}
%T.M. Rice, S. Gopalan and M. Sigrist, {\em Europhys. Lett.}, {\bf 23}, 
% 445 (1993).

\bibitem{Jerome91}
D.~J\'erome, {\em Science}, {\bf 252}, 1509 (1991).

\bibitem{Maekawa96}
S.~Maekawa, {\em Science}, {\bf 273}, 1515 (1996).

\bibitem{Osafune97}
T.~Osafune, N.~Motoyama, H.~Eisaki and S.~Uchida, {\em Phys. Rev. Lett.},
{\bf 78}, 1980 (1997).

\bibitem{Revco96}
U.~Ammerahl et al., to be published; A.~Revcolevschi and J.~Jegoudez,
Coherence in High Temperature Superconductors,
edited by G.~Deutscher and A.~Revcolevschi, p:428, World Scientific (1996).

\bibitem{Motoyama97}
N.~Motoyama, T.~Osafune, T.~Takeshita, H.~Eisaki and S.~Uchida, {\em Phys.
Rev. B}, {\bf 55}, R3386 (1997).

\bibitem{Akimitsu97}
J.~Akimitsu, T.~Nagata, M.~Uehara, J.~Goto, N.~Komiya, N.~Motoyama,
H.~Eisaki, S.~Uchida, H.~Takahashi, T.~Nakanishi and N.~M\^ori, preprint
(1997).

\bibitem{Benedek58}
G.B. Benedek and T.~Kushida, {\em J. Phys. Chem. Solids}, {\bf 5}, 241 (1958).

\bibitem{Caretta97}
P.~Caretta et~al., unpublished; M.~Takigawa, N.~Motoyama, H.~Eisaki and
S.~Uchida, preprint (1997).

\bibitem{Troyer94}
M.~Troyer, H.~Tsunetsugu and D.~W\"urtz, {\em Phys. Rev. B}, {\bf 50},
13515 (1994).

\bibitem{Magishi97}
K.~Magishi et~al., {\em Phys. Rev. B}, submitted (1997).

\bibitem{Kumagai97}
K.~Kumagai, S.~Tsuji, M.~Kato and Y.~Koike, {\em Phys. Rev. Lett.},{\bf
78}, 1992 (1997).

\bibitem{Orsay97}
P.~Auban-Senzier et~al., to be published.

\bibitem{Berthier96}
C.~Berthier, M.H. Julien, M.~Horvati\'c and Y.~Berthier, {\em J. Phys. I
(France)}, {\bf 6}, 2205 (1996).

\bibitem{Bourbonnais93}
C.~Bourbonnais, {\em J. Phys. I (France)}, {\bf 3},143 (1993).

\bibitem{Narath76}
A.~Narath, {\em Phys. Rev. B}, {\bf 13}, 3724 (1976).

\bibitem{weak-coupling}
H.J. Schulz, {\em Phys. Rev. B}, R2959 (1996); 
L.~Balents and M.P.A.~Fisher, {\em Phys. Rev. B}, 12133 (1996). 

\bibitem{TMRice}
T.~M\"uller and T.M.~Rice, {\em private communication}.

\bibitem{Riera97}
J.~Riera et~al., to be published. 

\end{references}
\end{document}